\documentstyle[prl,aps,psfig]{revtex}
\newcommand{\beq}{\begin{equation}}
\newcommand{\eeq}{\end{equation}}
\newcommand{\beqa}{\begin{eqnarray}}
\newcommand{\eeqa}{\end{eqnarray}}
\newcommand{\ba}{\begin{array}}
\newcommand{\ea}{\end{array}}

\begin{document}
\draft

\twocolumn[\hsize\textwidth\columnwidth\hsize\csname
@twocolumnfalse\endcsname

\widetext 

\title{Self-Trapping, Quantum Tunneling and Decay Rates \\ 
for a Bose Gas with Attractive Nonlocal Interaction} 
\author{Luca Salasnich} 
\address{Istituto Nazionale per la Fisica della Materia, Unit\`a di Milano,\\ 
Dipartimento di Fisica, Universit\`a di Milano, \\ 
Via Celoria 16, 20133 Milano, Italy} 

\maketitle

\begin{abstract} 
We study the Bose-Einstein condensation for a cloud of $^7$Li atoms 
with attractive nonlocal (finite-range) interaction in a harmonic trap. 
In addition to the low-density metastable branch, that is present 
also in the case of local interaction, a new stable branch 
appears at higher densities. 
For a large number of atoms, the size of the cloud in the stable 
high-density branch is independent of the trap size 
and the atoms are in a macroscopic quantum 
self-trapped configuration. 
We analyze the macroscopic quantum tunneling between the 
low-density metastable branch and the high-density one 
by using the istanton technique. Moreover we consider the 
decay rate of the Bose condensate 
due to inelastic two- and three-body collisions. 
\end{abstract}


]

\narrowtext

\par
The Bose-Einstein condensation (BEC) for atoms with effective 
attractive interaction, i.e. negative scattering length, 
is an interesting subject of study 
because of the instability of the system: 
Recent experiments with $^7$Li vapors in a harmonic trap 
have shown that the BEC is achieved only by a finite and 
small number of atoms [1]. In a uniform system, $^7$Li atoms do 
not form a stable BEC state because the s-wave 
scattering length $a_s$ is negative and the attractive 
interaction causes the condensate to collapse upon itself [2]. 
In a harmonic trap, atoms acquire zero-point 
energies which counterbalance the attractive interaction: 
a metastable condensate can be obtained 
for a number of atoms below a critical value $N_c$, which is 
about $10^3$ under the condition of present experiments [1,3]. 
\par 
Recently, we have shown that the inclusion of a nonlocal interaction 
implies the existence of a new stable branch intermediate 
in density between the known very dilute metastable state 
and the collapsed state [4]. 
In this paper, we discuss the properties of the stable state and 
calculate the tunneling rate of 
macroscopic quantum tunneling (MQT) from the 
metastable state to the stable one by using the istanton technique. 
Moreover, we determine the decay rate of the Bose condensate 
due to inelastic two- and three-body collisions. 
\par 
The system of $^7$Li atoms is quite dilute 
and at zero temperature the atoms are practically 
all in the condensate [1,3]. The action of the system is given by 
$$ 
S = \int dt \; d^3{\bf r} \; 
\phi^*({\bf r},t) \left[ i\hbar {\partial \over \partial t} 
+{\hbar^2 \over 2 m} \nabla^2 
- U({\bf r}) \right] \phi({\bf r},t) 
$$
\beq 
-{1\over 2} \int dt \; d^3{\bf r} \; d^3{\bf r}' \; \phi^*({\bf r},t) 
\phi^*({\bf r}',t) V(|{\bf r}-{\bf r}'|) 
\phi({\bf r}',t) \phi({\bf r},t) \; , 
\eeq 
where $\phi ({\bf r},t)$ is the macroscopic wavefunction 
of the condensed atoms, $U({\bf r})$ is the external potential and 
$V(|{\bf r}-{\bf r}'|)$ is the interatomic potential [5]. 
\par
For alkali atoms, the atom-atom interaction is usually replaced 
by an effective zero-range pseudo-potential, 
$V(r)=g\delta^3(r)$, where $g={4\pi \hbar^2 a_s/m}$ 
is the scattering amplitude and $a_s$ is the s-wave scattering length. 
By using such an effective potential, and imposing 
the least action principle to Eq. (1), one gets 
the so-called Gross-Pitaevskii (GP) equation [6]. 
\par
In the case of $^{7}Li$, that has negative scattering length, 
the scattering cross section has a significant 
momentum dependence already at very low momenta [7,8]. 
This momentum dependence implies that the effective interaction is non local 
changing from attractive to repulsive at a characteristic range $r_e$. 
We assume [4] that the attractive potential has a finite range $r_e$ and
in addition we allow for the presence of a repulsive contribution
which is modeled as a {\it local} positive term defined by a "high energy"
scattering length $a_r > 0$. The effective interaction can be written 
in the following form: 
\beq 
V(k)={4\pi\hbar^2\over m}\left[a_r + (a_s-a_r) f(kr_e) \right] \; . 
\eeq 
We have considered two choices for the shape function f(x): 
a Lorenzian $f(x)=(1+x^2)^{-1}$ and a Gaussian 
$f(x)=\exp{(-x^2)}$. The results do not depend on the specific choice 
of $f(x)$ and so we will discuss only the Lorenzian case.  
We use interaction parameters appropriate for $^7Li$ 
(see Parola, Salasnich and Reatto [4]): 
$a_s=-27\,a_B$, $r_e=10^3\,a_B$ and $a_r=6.6\,a_B$, 
where $a_B$ is the Bohr radius. 
\par 
To study the properties of the Bose condensate we perform a 
variational calculation. Because of the external trap, that we model with an 
isotropic potential $U({\bf r})=(1/2)m\omega_0^2 r^2$, 
we use a Gaussian trial wavefunction to 
describe the Bose condensate and minimize the quantum action. 
In the rest of the paper we write the lengths in units 
of the characteristic length of the trap $a_0=\sqrt{\hbar /(m\omega_0)}$, 
the time in units $\omega_0^{-1}$, the action in units $\hbar$ and 
the energy in units $\hbar \omega_0$. A good choice for the 
wavefunction is the following 
\beq 
\phi({\bf r},t)= N^{1/2} {1\over \pi^{3/4} {\sigma}^{3/2}(t)}  
\exp\left\{-{r^2\over 2\sigma^2(t)}
+i \beta(t) r^2 \right\} \; ,
\eeq 
where $N$ is the number 
of condensed atoms and $\sigma$ and $\beta$ are 
time-dependent variational parameters. $\sigma$ is 
the width of the condensate. Note that, in order to describe 
the time evolution of the variational 
function, the phase factor $i \beta(t) r^2$ is needed [9]. 
The choice of a Gaussian shape for the condensate 
is well justified in the limit of weak interatomic coupling, 
because the exact ground-state of the linear Schr\"odinger 
equation with harmonic potential is a Gaussian. Moreover, 
for the description of the collective 
dynamics of Bose-Einstein condensates, 
it has already been shown that the variational technique based 
on Gaussian trial functions leads to 
reliable results even in the large condensate number limit [9]. 
By inserting the trial wavefunction in Eq. (1), 
after spatial integration, the action of the system reads 
\beq 
S= {N\over 4} \int dt \; \left( 3 {\dot \sigma}^2 - W(\sigma) \right) \; , 
\eeq 
where the effective potential energy is given by 
\beq 
W(\sigma) =  {3\over \sigma^2}+3\sigma^2+2 N 
\left( {\gamma_r \over \sigma^3} + 
{\tau_1\over \sigma} - \tau_2 g(\chi^2\sigma^2) 
\right) \; ,
\eeq
with $\gamma_r=(2/\pi)^{1/2}a_r/a_0$, 
$\tau_1=(2/\pi)^{1/2}a_0(a_s-a_r)/r_e^2$, 
$\chi=2^{-1/2}a_0/r_e$, $\tau_2=a_0^2(a_s-a_r)/r_e^3$ 
and $g(x)= \exp{(x)}(1-{\rm erf}(x))$, where ${\rm erf}(x)$ is 
the error function. Note that the action does not depend explicitly 
on $\beta$, which is fully determined by $\sigma$ via 
the Euler-Lagrange equation $\beta = -{\dot \sigma}/(2 \sigma)$ 
(for further details see [10]). 
\par
The extrema of $W(\sigma )$ are obtained 
as solutions of an algebraic equation, which 
gives the number of bosons as a function of the size 
$\sigma$ of the cloud 
\beq 
N =(1-\sigma^4)
\left[ -\gamma_r \sigma^{-1} -{\tau_1 \sigma\over 3} 
+{2\chi\tau_2\sigma^3\over 3 \sqrt{\pi}} -{2\chi^2\tau_2 
\sigma^4 g(\chi \sigma )\over 3} \right]^{-1} \; . 
\eeq
This equation has either one or three positive roots 
depending on the parameters and on number $N$ of atoms in the cloud. 
When three solutions are present, the intermediate one represents 
an unstable state (i.e. a local maximum of the energy), 
while the other two respectively describe a low-density metastable 
solution (local minimum) and a high-density stable solution (absolute minimum) 
within GP approximation. In the local case, Eq. (6) reduces 
to $N=(\sigma^5 - \sigma)/\gamma$, where $\gamma=(2/\pi)^{1/2}|a_s|/a_0$. 
\par
The variational results for four trap
sizes are shown in Fig. 1, where the size of the condensate 
is plotted as a function of $N$ 
(see also Reatto, Parola and Salasnich [4]). 
For comparison, we also show the radius of the cloud when a local 
interaction is assumed. In this case there 
is a critical number $N_c\simeq 0.67 a_0/a_s$ of bosons beyond 
which the cloud collapses [3]. 
Fig. 1 shows that the effects of non-locality are always 
important for very tight traps 
while for larger traps non-locality is important just when the radius of 
the cloud rapidly drops for increasing $N$. 
This "transition" is discontinuous for large traps. 
By reducing the trap size, however, this 
discontinuity is strongly reduced and, below about $a_0=0.3\,\mu$m, 
the unstable branch disappears and there is a smooth evolution from a very 
dilute cloud to a less dilute state with an increasing density as $N$ grows.  
For large $N$ the size of the cloud is independent of the trap size 
and the atoms are in a macroscopic quantum 
self-trapped (MQST) configuration. 
Note that we have also computed the exact 
solution of the GP equation. 
The variational approach is always very close to 
the exact solution [4]. 
\par
The condensate undergoes density oscillations around the 
minimum of the energy $W(\sigma )$. 
By performing a quadratic Taylor expansion around the minimum 
one finds simple expressions for the monopole collective frequency. 
In the local case we have $\omega = (5-\sigma^{-4})^{1/2}$, 
such that $\omega = 2$ for $\sigma=1$ (ideal gas), 
and $\omega \to 0$ for $\sigma_0\to 5^{-1/4}$ (i.e. for $N\to N_c$). 
Instead, in the nonlocal case we obtain 
$$
\omega  = \left[ 3\sigma^{-4}+1+N 
\left( {4 \gamma_r \over \sigma^{5}} + 
{2\tau_1 \over 3\sigma^3} -{2\chi^2\tau_2 g(\chi \sigma)
(1+2\chi^2\sigma^2) \over 3} \right. \right. 
$$
\beq
\left. \left. + {4\sigma \chi^3 \tau_2\over 3\sqrt{\pi}} 
\right) \right]^{1/2} \; ,
\eeq 
where $\sigma$ is related to $N$ by Eq. (6). 
For the larger traps, where the two branches are present, 
the frequency of the metastable branch is very close to the result given by 
the local approximation. In the stable branch $\omega$ starts from zero 
and it raises rapidly as $N^{1/2}$. 
For traps of intermediate size, 
$\omega$ has a dip in the transition region between 
the low-density state and the self bound state. For very small traps, 
there is only one branch and the frequency increases smoothly with the number 
of bosons. Note that the variational monopole frequency 
and the numerical one (non local case), 
obtained by solving the nonlocal time-dependent GP equation, 
differ by less than $3\%$ (see Reatto, Parola and Salasnich [4]). 
\par 
We have seen that the low-density and high-density 
branches represent local minima 
of the effective potential energy $W(\sigma )$ of the condensate,  
given by Eq. (5). We know from quantum mechanics 
that a state that is concentrated in one minimum of the potential energy 
may tunnel into a lower one. 
Ueda and Leggett [11] have shown that in the local 
case, very close to $N_c$, there is macroscopic quantum tunneling (MQT) 
from the metastable branch to the collapsed state, which has 
zero radius and energy equal to minus infinity. 
In our more realistic picture, there is MQT from the low-density 
branch to the high-density one. 
The rate of MQT can be calculated by the semiclassical formula 
$\Gamma_T = A e^{-{\tilde S}/\hbar}$, where 
${\tilde S}$ is the istantonic action, that is 
the action $S$ with imaginary-time (Euclidean action)  
evaluated along the trajectory that makes it extremal [12]. 
For a quadratic-plus-cubic potential 
the prefactor $A$ is given by $A=\omega (15 S^B/2\pi \hbar )^{1/2}$ [11]. 
After some tedious but straightforward calculations, one finds 
\beq 
\Gamma_T = 36\sqrt{3\over 2 \pi} {\omega^{7/2} 
N^{1/2} \over {d^3W\over d\sigma^3} } \exp{\left\{ -{36^2\over 5}
{N\omega^5 \over ({d^3W\over d\sigma^3})^2} \right\}} \; , 
\eeq 
where $W$, $N$ and $\omega$ are given by Eq. (5), (6) and (7), respectively. 
In Fig. 2 we plot the rate of MQT for four different traps. 
As expected, this rate becomes large only near $N_c$, both in the 
local and nonlocal case. Moreover, $\Gamma_T$ increases strongly by reducing 
the trap size $a_0$. As previously discussed, 
in the nonlocal case, below about $a_0=0.3\,\mu$m 
the unstable branch disappears and there is no more MQT. 
\par 
We observe that, also if the high density branch is mechanically stable, 
it has a very short life-time due to two-body dipolar collisions 
and three-body recombination. Atoms that inelastically collide acquire 
substantial energy and are ejected from the trap. 
The total loss rate due to two-body 
dipolar collision and three-body recombination collisions is given by 
$\Gamma_L = K \int d^3{\bf r} |\phi({\bf r})|^4 + 
L \int d^3{\bf r} |\phi({\bf r})|^6$, 
where $K=1.2\cdot 10^{-2}$ $\mu$m$^3$ sec$^{-1}$ is the two-body 
coefficient and $L=2.6\cdot 10^{-4}$ $\mu$m$^6$ sec$^{-1}$ the 
three-body coefficient [13]. By using the trial wavefunction of Eq. (3), 
we find 
\beq
\Gamma_L={K\over (2\pi)^{3/2} a_0^3} {N^2\over \sigma^3} + 
{L\over (3\pi^2)^{3/2} a_0^6} {N^3\over \sigma^6} \; . 
\eeq 
In this formula the value of $N$ is related to that of $\sigma$ 
through Eq. (6). In Fig. 3 we plot the total loss rate $\Gamma_L$ 
as a function of the number of $^7$Li atoms. 
During MQT the condensate density grows rapidly, thereby increasing 
the decay rate from inelastic two- and three-body collisions. 
In the case of a unique stable branch, 
the loss rate increases very fast with the number of atoms 
due to the high-density of the system. 
For sake of completeness, in Fig. 4 we show 
the condensate size, the MQT rate and 
the loss rate for the trap of 
the Rice experiment ($a_0=3$ $\mu$m). Actually, the trap 
of the Rice group has a small anisotropy and we have 
used the geometric average of the frequencies [1,4]. 
It is interesting to observe that for this trap 
the high density branch exists also 
for a number of atoms much smaller than $N_c\simeq 1400$. 
\par 
In conclusion, we have studied the effect of a nonlocal inter-atomic 
interaction for a Bose-Einstein condensate of $^7$Li atoms 
in a isotropic harmonic potential at zero temperature. 
We have shown that, 
in addition to the low-density metastable branch, which is present 
also in the case of local interaction, a new stable branch 
appears at higher densities. 
For a large number of atoms, the condensate is in a novel 
macroscopic quantum self-trapped (MQST) state. 
Very close to the low-density critical threshold, 
there is macroscopic quantum tunneling (MQT) between this 
low-density metastable branch and the high-density one. 
This prediction is more realistic than that of MQT 
from the metastable branch to a collapsed state, which has 
zero radius and energy equal to minus infinity. 
We have calculated the rate of MQT with the semiclassical 
technique of istantons. 
Finally, we have considered the decay rate of the Bose condensate 
due to inelastic two-body dipolar collisions 
and three-body recombination. The life-time of the condensate 
in the high-density branch is very short because of the small 
radius of the condensate in these configurations. 
Due to the very short life-time of $^7$Li, experimental studies 
of the high-density phase of the condensate are practically impossible. 
Nevertheless, this is only a technical difficulty, 
perhaps surmountable by using a different atom, and it is imaginable 
that in the future our predictions can be checked. 
\par
The author is grateful to A. Parola and L. Reatto 
for stimulating discussions. This work has been supported by INFM 
under the Research Advanced Project (PRA) 
on "Bose-Einstein Condensation". 

\section*{References}

\begin{description}

\item{\ [1]} C.C. Bradley, C.A. Sackett, J.J. Tollett, and R.G. Hulet, 
Phys. Rev. Lett. {\bf 75}, 1687 (1995); 
C.C. Bradley, C.A. Sackett, and R.G. Hulet, 
Phys. Rev. Lett. {\bf 78}, 985 (1997); 
C.A. Sackett, H.T.C. Stoof, and R.G. Hulet, 
Phys. Rev. Lett. {\bf 80}, 2031 (1998). 

\item{\ [2]} T.D. Lee, K. Huang, and C.N. Yang, 
Phys. Rev. {\bf 106}, 1135 (1957). 

\item{\ [3]} R.J. Dodd, M. Edwards, C.J. Williams, C.W. Clark, 
M.J. Holland, P.A. Ruprecht and K. Burnett, Phys. Rev. A {\bf 54}, 661 (1996); 
G. Baym and C.J. Pethick, Phys. Rev. A {\bf 54}, 6 (1996). 

\item{\ [4]} A. Parola, L. Salasnich and L. Reatto, 
Phys. Rev. A {\bf 57}, R3180 (1998); 
L. Reatto, A. Parola, and L. Salasnich, 
J. Low Temp. Phys. {\bf 113}, 195 (1998); 
L. Salasnich, Mod. Phys. Lett. B {\bf 12}, 649 (1998). 

\item{\ [5]} A.L. Fetter and J.D. Walecka, {\it Quantum Theory 
of Many-Particle Systems} (McGraw-Hill, New York, 1971); 
K. Huang, {\it Statistical Mechanics} (Wiley, New York, 1963). 

\item{\ [6]} E.P. Gross, Nuovo Cimento {\bf 20}, 454 (1961); 
J. Math. Phys. {\bf 4}, 195 (1963); 
L.P. Pitaevskii, Zh. Eksp. Teor. Fiz. {\bf 40}, 646 (1961) 
[Sov. Phys. JETP {\bf 13}, 451 (1961)].

\item{\ [7]} G.F. Gribakin and V.V. Flambaum, Phys. Rev. A {\bf 48}, 
546 (1993). 

\item{\ [8]} R. Cot\`e, A. Dalgarno and M.J. Jamieson, 
Phys. Rev. A {\bf 50}, 399 (1994).  

\item{\ [9]} E. Cerboneschi, R. Mannella, E. Arimondo, L. Salasnich, 
Phys. Lett. A {\bf 249}, 245 (1998). 

\item{\ [10]} L. Salasnich, "Time-Dependent Variational Approach 
to Bose-Einstein Condensation", cond-mat/9908147, to be published 
in Int. J. Mod. Phys. B. 

\item{\ [11]} M. Ueda and A.J. Leggett, Phys. Rev. Lett. 
{\bf 80}, 1576 (1998). 

\item{\ [12]} M. Kaku, {\it Quantum Field Theory: a Modern Introduction} 
(Oxford Univ. Press, Oxford, 1993). 

\item{\ [13]} A.J. Moerdijk, H.M.J.M.Boesten, and B.J. Verhaar, 
Phys. Rev. A {\bf 53}, 916 (1996); 
H. Shi and W.M. Zheng, Phys. Rev. A {\bf 55}, 2930 (1997);  
C.A. Sackett, J.M. Gerton, M. Welling, and R.G. Hulet, 
Phys. Rev. Lett. {\bf 82}, 876 (1999). 

\end{description}

\begin{figure}
\centerline{\psfig{file=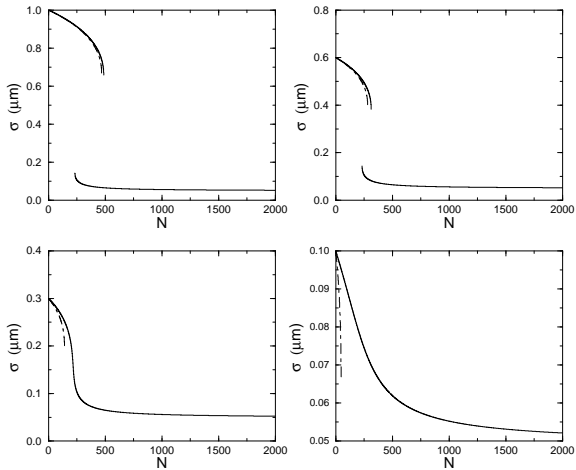,height=2.5in}}
\caption{Size of the condensate 
as a function of the number $N$ of $^7$Li atoms. 
Four different traps, from top to bottom and 
from left to right: $a_0=1.0$ $\mu$m, $0.6$ $\mu$m, 
$0.3$ $\mu$m, $0.1$ $\mu$m, respectively.  
The dashed lines represent the results with local interaction.}  
\end{figure}

\begin{figure}
\centerline{\psfig{file=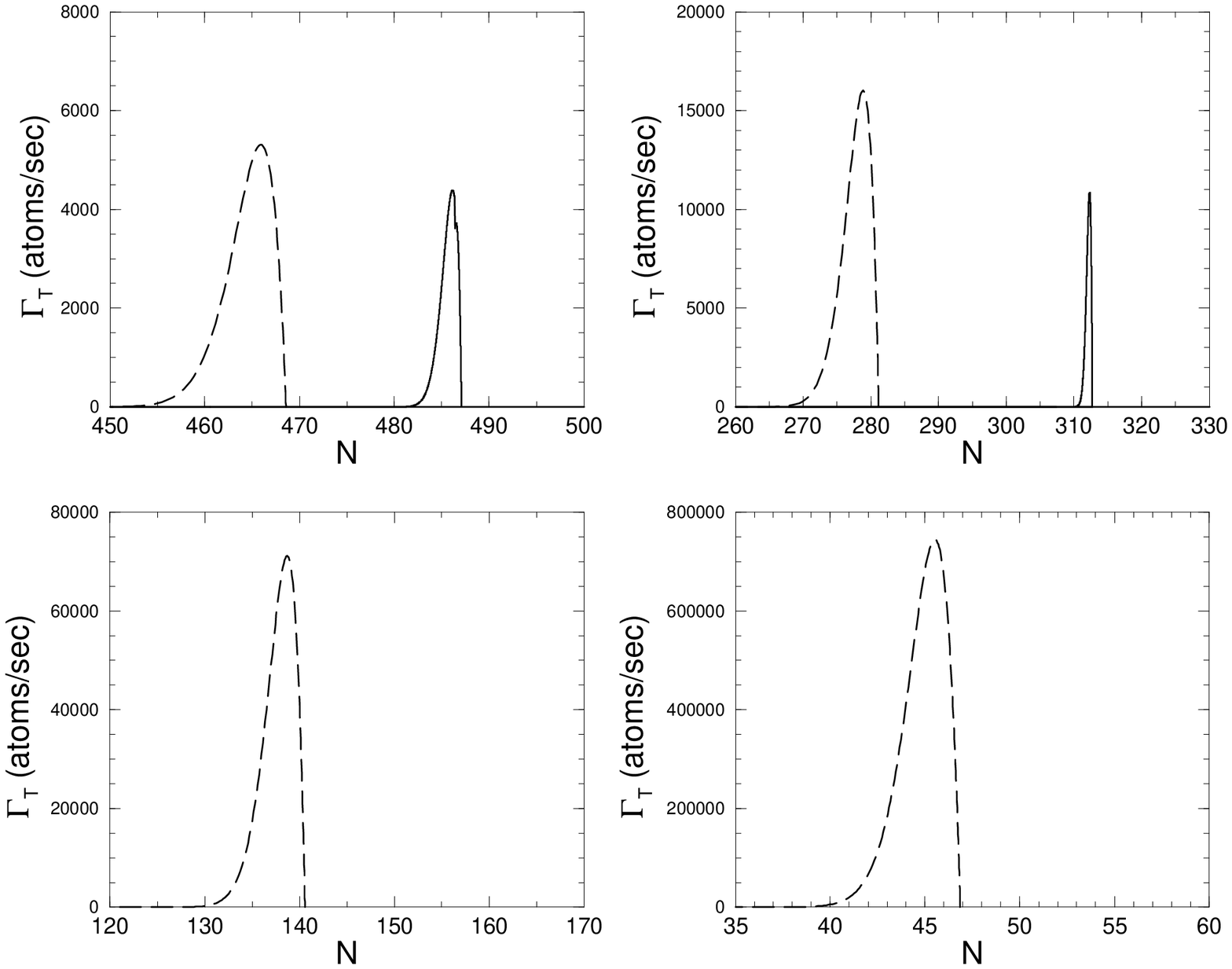,height=2.5in}}
\caption{Rate of MQT as a function of the number $N$ 
of $^7$Li atoms. Four different traps, from top to bottom and 
from left to right: $a_0=1.0$ $\mu$m, $0.6$ $\mu$m, $0.3$ $\mu$m, 
$0.1$ $\mu$m, respectively. 
The dashed lines represent the results with local interaction.}  
\end{figure}

\begin{figure}
\centerline{\psfig{file=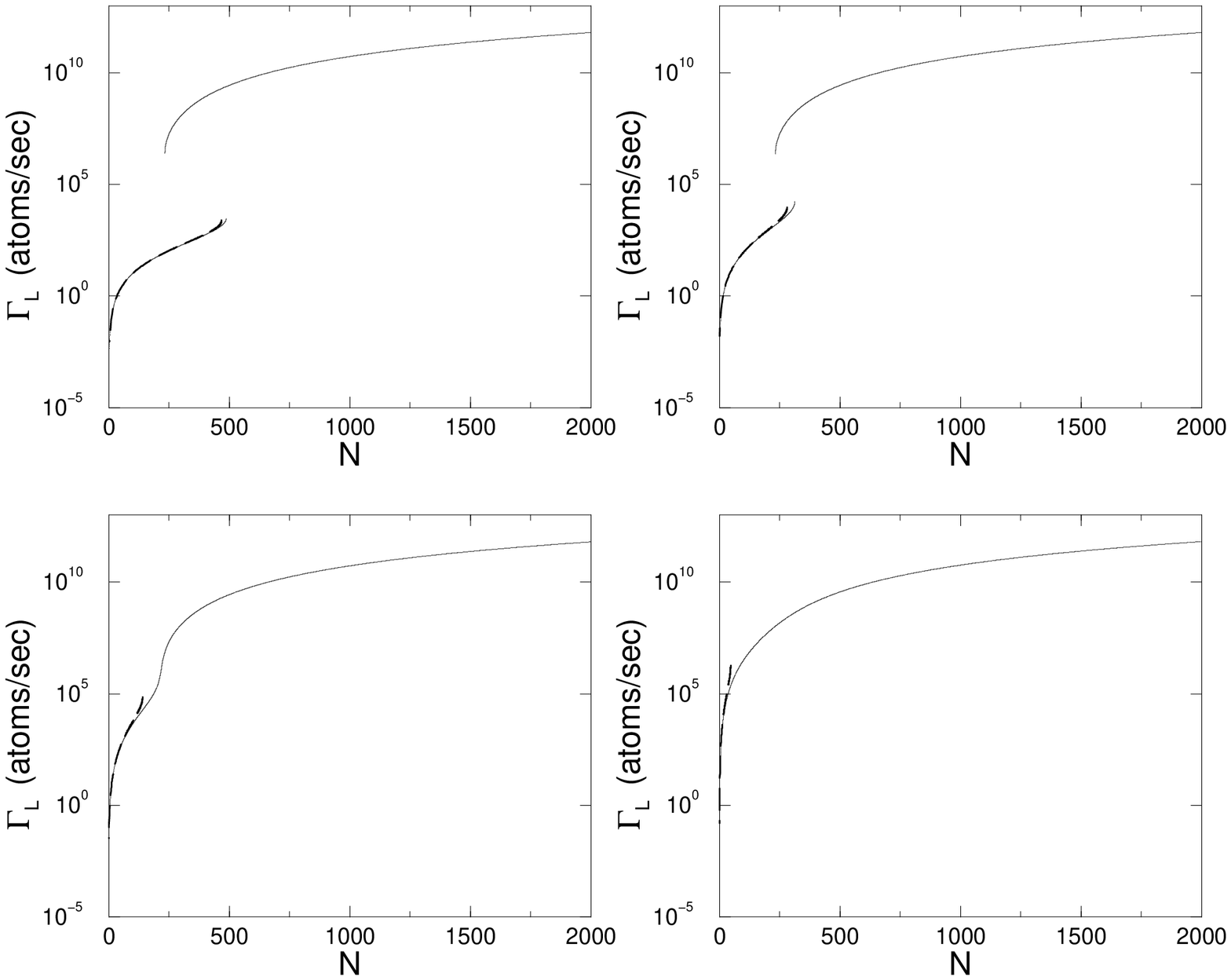,height=2.5in}}
\caption{Loss rate due to two- and three-body collisions 
as a function of the number $N$ of $^7$Li atoms. 
Four different traps, from top to bottom and 
from left to right: $a_0=1.0$ $\mu$m, $0.6$ $\mu$m, $0.3$ 
$\mu$m, $0.1$ $\mu$m, respectively. 
The dashed lines represent the results with local interaction.}  
\end{figure}

\begin{figure}
\centerline{\psfig{file=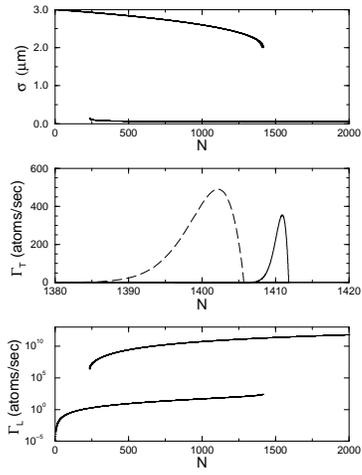,height=2.5in}}
\caption{Condensate size $\sigma$, MQT rate $\Gamma_T$ 
and loss rate $\Gamma_L$ as a function 
of the number $N$ of $^7$Li atoms. Trap of Rice University 
with $a_0=3.0$ $\mu$m [1]. 
The dashed lines represent the results with local interaction.}  
\end{figure}

\end{document}